\documentclass[aps,prd,reprint,groupedaddress,nofootinbib]{revtex4-1}
\usepackage{graphicx}
\usepackage{color}
\usepackage{ulem}

\begin{document}


\title{Critical exponents of nonequilibrium phase transitions in AdS/CFT correspondence}


\author{Masataka Matsumoto and Shin Nakamura}
\affiliation{Department of Physics, Chuo University, Tokyo 112-8551, Japan}



\begin{abstract}
We study critical phenomena of nonequilibrium phase transitions by using the AdS/CFT correspondence. Our system consists of charged particles interacting with a heat bath of neutral gauge particles. The system is in current-driven nonequilibrium steady state, and the nonequilibrium phase transition is associated with nonlinear electric conductivity. We define a susceptibility as a response of the system to the current variation. We further define a critical exponent from the power-law divergence of the susceptibility. We find that the critical exponent and the critical amplitude ratio of the susceptibility agree with those of the Landau theory of equilibrium phase transitions, if we identify the current as the external field in the Landau theory. 
\end{abstract}


\maketitle


\section{Introduction}
\label{Introduction}
Nonequilibrium phenomena have wider variety compared to equilibrium phenomena. 
The number of parameters that control the nonequilibrium systems is larger than that of the equilibrium systems, in general. 
In nonequilibrium steady states (NESSs), the typical additional parameter that is absent from equilibrium systems is {\it current}. For example, a system attached to two heat baths of different temperatures has a heat current. In this case, the heat current is the ``new'' parameter, which is special to nonequilibrium systems.
Another example is an electric current along an electric field in a conductor. In this case, the equilibrium state is realized when the current is vanishing.

These new parameters, the heat current and the electric current, measure the rate of entropy production (when the electric field or the temperatures of two heat baths are kept fixed), hence represent the ``distance'' from the equilibrium states. A primary question in nonequilibrium physics is how these parameters control the systems. In particular, investigation of phase transitions under the presence of current is one important research subject.\footnote{A first-order phase transition in the presence of {\it heat current} has been studied in \cite{Nakagawa-Sasa}.}

In this paper, we study nonequilibrium critical phenomena driven by an electric current density $J$. When a system exhibits a second order phase transition under the presence of $J$, a natural question is how the new parameter $J$ controls the critical phenomena. More specifically, the following questions can be addressed. 1) Is it possible to define a susceptibility with respect to $J$ in a sensible way?  2) If yes, how does it behave near the critical point? Are there any critical phenomena associated with the new parameter $J$? If it is the case, what are the critical exponents? 3) Do we have more critical exponents for  nonequilibrium phase transitions than what we have for equilibrium systems? Can we construct a theory for these critical exponents?

In order to reveal these issues, we employ the anti-de Sitter/conformal field theory (AdS/CFT) correspondence.
The AdS/CFT correspondence is a duality between a classical gravity theory and a strongly coupled quantum gauge field theory~\cite{Maldacena,GKP}. This correspondence provides a computational method for the gauge field theory beyond the linear response regime in terms of general relativity.\footnote{A review on application of the AdS/CFT correspondence to nonequilibrium physics is Ref.~\cite{Hubeny:2010ry}.} 

For example, a NESS of strongly interacting gauge theory plasma was studied in Ref.~\cite{SN_PTP}. It was shown in Ref.~\cite{SN_PTP} that the system exhibits not only positive differential conductivity (PDC) but also negative differential conductivity (NDC) in the NESS driven by a constant current. Furthermore, a first-order and a second-order nonequilibrium phase transitions associated with the nonlinear conductivity were discovered in the same system \cite{SN_PRL}.\footnote{A same type of phase transition was also observed later in a different setup of holography in Ref.~\cite{AA}.}
In Ref.~\cite{SN_PRL}, critical exponents $\beta$ and $\delta$ for the nonequilibrium phase transition were defined.\footnote{The critical exponent $\delta$ in this paper is $\tilde{\delta}$ in Ref.~\cite{SN_PRL}.} 
In the Landau theory of equilibrium phase transitions, critical exponent $\delta$ is defined from the power-law dependence of the order parameter with respect to an external field (e.g. a magnetic field in ferromagnets). In this case, the external field does not drive the system into nonequilibrium states.
On the other hand, the exponent $\delta$ in the nonequilibrium phase transition was defined as the exponent of the power behavior of the order parameter in the variation of the current. 
Interestingly, the obtained values of the $\beta$ and the $\delta$ in the nonequilibrium phase transition agreed with those of $\beta$ and $\delta$ in the Landau theory of equilibrium phase transitions, respectively, within the numerical error~\cite{SN_PRL}. This result implies that a current density $J$, which is a parameter that appears only in the nonequilibrium system, plays a fundamental role in characterizing the nonequilibrium phase transition. However, we still lack a complete answer to the questions 1), 2) and 3) raised above.
 
In this paper, we further proceed calculations of critical exponents associated with $J$ in the nonequilibrium phase transition. We define a susceptibility as a response of the system to the current variation. We also define a critical exponent $\gamma$ from the power-law divergence of the susceptibility. 
We find that the susceptibility shows critical phenomena and the value of the critical exponent $\gamma$ agrees with that in the Landau theory of equilibrium phase transitions.
The critical amplitude ratio of the susceptibility also agrees with that in the Landau theory.
Together with the results for $\beta$ and $\delta$, our results state that the critical phenomena in the nonequilibrium phase transition in question have remarkable similarity with those in the Landau theory of equilibrium phase transitions, if we identify the current as the external field.\footnote{In the present paper, we employ the same notations $\alpha$, $\beta$, $\delta$ and $\gamma$ for the critical exponents of our nonequilibrium phase transition as those of the Landau theory given in Sec.~\ref{LandauTheory}. However, their physical definitions should be distinguished.}

The organization of the paper is as follows. In Sec.~\ref{LandauTheory}, we review the Landau theory of equilibrium phase transitions. In Sec.~\ref{Setup}, we explain the setup of our model. We focus on the so-called D3-D7 system. In Sec.~\ref{CriticalExponents}, we propose the definitions of the susceptibility and the critical exponent $\gamma$ mentioned above. We compute them and the results will be exhibited. We conclude in Sec.~\ref{Conclusion}.

\section{Landau Theory and Critical Exponent}
\label{LandauTheory}
We begin with a brief review of the Landau theory of equilibrium phase transitions. The new definitions of the susceptibility and the critical exponents in Sec.~\ref{CriticalExponents} will be given by using an analogy with the Landau theory.

Here we consider a phase transition of ferromagnets. In the Landau theory, we assume the free energy is given as an even function of the magnetization because of the symmetry under the spin flip. Since we are interested in the critical point,  we assume the magnetization is sufficiently small. Then we can expand the free energy as a power series in the magnetization and ignore the higher order terms:
\begin{equation}
    F(M)=F_0 + a M^2 + bM^4,
    \label{LandauFreeEnergy}
\end{equation}
where $M$ is the magnetization as an order parameter. $F_0$ and $b$ are constants, whereas
$a=k(T-T_c)/T_c$ with a constant $k$. $T_c$ is the critical temperature of the second-order phase transition.

Since a thermal equilibrium state is realized as a minimum of the free energy,  we consider
\begin{equation}
    \left. \frac{\partial F(M)}{\partial M} \right|_{M_0}=2 a M_0 + 4 b M_0^3 =0,
    \label{FreeEnergyDiff}
\end{equation}
to obtain the thermal equilibrium magnetization $M_0$:
\begin{equation}
    M_0=\sqrtsign{-\frac{a}{2b}}=\sqrtsign{\frac{k(T_c-T)}{2bT_c}},
\label{m0}
\end{equation}
for $T<T_c$.
If there is an external magnetic field in this system, the free energy is written as
\begin{equation}
    F(M)=F_0 + a M^2 + bM^4 -HM,
    \label{LandauFreeEnergyH}
\end{equation}
where $H$ is the external magnetic field. Then, the thermal equilibrium state is determined by the following relation:
\begin{equation}
     \left. \frac{\partial F(M)}{\partial M} \right|_{M_0}=2 a M_0 + 4 b M_0^3 -H =0.
     \label{FreeEnergyDiffH}
\end{equation}
Although the solution of Eq.~(\ref{FreeEnergyDiffH}) is complicated, $M_0$ becomes simple for $T=T_c$:
\begin{equation}
    M_0=\left( \frac{H}{4b} \right) ^{\frac{1}{3}}.
    \label{m0H}
\end{equation}

The magnetic susceptibility is defined as $\chi=\partial M / \partial H$. We obtain this from Eq.~(\ref{FreeEnergyDiffH}) as
\begin{equation}
    \chi=\frac{1}{2a+12b M^2}.
    \label{ChiDef}
\end{equation}
For $T>T_c$ with $M=0$, we have
\begin{equation}
    \chi=\frac{1}{2a}=\frac{T_c}{2k(T-T_c)} \equiv \chi_{_{ T>T_c}},
    \label{Chi1}
\end{equation}
whereas
\begin{equation}
    \chi=\frac{1}{2a+12b(-a/2b)}=\frac{T_c}{4k(T_c-T)} \equiv \chi_{_{T<T_c}},
    \label{Chi2}
\end{equation}
for $T<T_c$. From Eqs.~(\ref{Chi1}) and~(\ref{Chi2}), the ratio of the coefficients of $|T-T_c|$, which is called as {\it critical amplitude ratio}, becomes two: $\chi_{_{T>T_c}} / \chi_{_{T<T_c}}=2$. Note that this value is independent of $k$ and $T_c$ we have introduced.

Let us derive the specific heat from the free energy. The specific heat is defined by $C_v=-T \partial^2 F(M_0) / \partial T^2$. For $T<T_c$ and $H=0$,
\begin{equation}
     F(M_0)=F_0 + a M_0^2 + bM_0^4 =F_0 - \frac{k^2(T-T_c)^2}{4bT_c^2}.
     \label{SpecoficHeatCapacity}
\end{equation}
Thus we find that the specific heat is constant.

The divergent behaviors of various quantities at the critical point are characterized by the {\it critical exponents}. The definitions of the critical exponents in ferromagnets are given by
\begin{eqnarray}
    M_0 \propto |T-T_c|^{\beta} \:\: (T<T_c), \label{CEbeta} \\
    M_0 \propto |H|^{1/\delta} \:\: (T=T_c), \label{CEdelta} \\
    \chi \propto |T-T_c|^{-\gamma} \:\: (T<T_c), \\
    \chi \propto |T-T_c|^{-\gamma'} \:\: (T>T_c), \\
    C_v \propto |T-T_c|^{-\alpha} \:\: (T<T_c).
\end{eqnarray}
In the Landau theory, the critical exponents are determined by Eqs.~(\ref{m0}), (\ref{m0H}), (\ref{Chi1}), (\ref{Chi2}), and (\ref{SpecoficHeatCapacity}) as:
\begin{equation}
    \beta=\frac{1}{2}, \:\:\:\: \delta=3, \:\:\:\: \gamma=\gamma'=1, \:\:\:\: \alpha=0,
    \label{MFCriticalExponents}
\end{equation}
and these values are the same as those in the mean-field theory. Note that there are further two critical exponents $\eta$ and $\nu$, which are related to the correlation functions. However, these critical exponents cannot be determined within the above discussion. We are not going to deal with $\eta$ and $\nu$ in the present paper.

\section{Setup}
\label{Setup}
To realize a system in NESS, we employ (3+1)-dimensional $SU(N_c)$ $\mathcal{N}=4$ super-symmetric Yang-Mills (SYM) theory with a fundamental $\mathcal{N}=2$ hypermultiplet as the microscopic theory. The theory contains the gauge particles in the adjoint representation (which we call gluon sector) and the charged particles (quark sector) in the fundamental and antifundamental representation. Here the charge is that of the global $U(1)_B$ symmetry, and not that of the color. In this sense, the gluon sector is neutral. We apply an constant external electric field acting on this charge. We take the large-$N_c$ limit in order to realize a NESS. This is because the degree of freedom of the gluon sector, which is $\mathcal{O}(N_c^2)$, becomes sufficiently larger than that of the quark sector, which is $\mathcal{O}(N_c)$. Then we can ignore the backreaction to the gluon sector in this limit. As a result, the gluon sector acts as a heat bath for the quark sector. Then the system realizes a NESS with a constant current of the charge.
The D3-D7 system is the gravity dual of our microscopic theory~\cite{KK_JHEP}. The D7-brane is embedded in the background geometry which is a direct product of a 5-dimensional AdS-Schwarzschild black hole (AdS-BH) and $S^5$. The gluon sector and the quark sector correspond to AdS-BH and the D7-brane, respectively. 

The metric of the AdS-BH part is given by
\begin{equation}
    ds^2=-\frac{1}{z^2}\frac{(1-z^4/z_{H}^{4})^2}{1+z^4/z_{H}^4}dt^2+\frac{1+z^4/z_{H}^{4}}{z^2}d\vec{x}^2+\frac{dz^2}{z^2},
    \label{AdSBH}
\end{equation}
where $z$ ($0 \leq z \leq z_H $) is the radial coordinate of the geometry. The boundary is located at $z=0$, whereas the horizon is located at $z=z_H$. The Hawking temperature is given by $T=\sqrt{2}/(\pi z_H)$. $t$ and $\vec{x}$ denote the (3+1)-dimensional spacetime coordinates of the gauge theory. The metric of the $S^5$ part is given by
\begin{equation}
    d\Omega_{5}^2=d\theta^2+\sin^2\theta d\psi^2+\cos^2\theta d\Omega_{3}^2,
    \label{S5}
\end{equation}
where $0 \leq \theta \leq \pi/2$ and $d\Omega_{d}$ is the volume element of a $d$-dimensional unit sphere. For simplicity, the radius of the $S^5$ part has been taken to be 1. This is equivalent to choosing the 't Hooft coupling $\lambda$ of the gauge theory at $\lambda=(2\pi)^2/2$.

In our D3-D7 system, the D7-brane is wrapped on the $S^3$ part of the $S^5$. Since the radius of the $S^3$ part is $\cos\theta$, the configuration of the D7-brane is determined by the function $\theta(z)$. The asymptotic form of $\theta(z)$ is given by
\begin{equation}
    \theta(z) = m_q z + \frac{1}{2} \left( \frac{ \left< \bar{q} q \right> }{N}+\frac{m_{q}^3}{3} \right) z^3 + \mathcal{O}(z^5),
    \label{theta}
\end{equation}
where $\left< \bar{q} q \right>$ denotes the chiral condensate and $m_q$ is the current quark mass~\cite{Babington:2003vm,Kruczenski:2003uq}. (See also Ref.~\cite{KO_JHEP}.) $N= T_{D7} (2\pi^2)=N_c  /(2 \pi)^2$ in our convention.

The D7-brane action is given by the Dirac-Born-Infeld (DBI) action:
\begin{equation}
    S_{D7}=- T_{D7} \int d^8 \xi \sqrt{-\det(g_{ab}+(2\pi\alpha')F_{ab})} .
    \label{DBIaction}
\end{equation}
Here $T_{D7}$ is the D7-brane tension, $\xi$ are the world-volume coordinates, $g_{ab}$ is the induced metric and $F_{ab}$ is the $U(1)$ field strength on the D7-brane. The Wess-Zumino term does not contribute in our setup. Assuming the external electric field $E$ is applied along the $x$ direction, the asymptotic form of the gauge field $A_{x}$ on the D7-brane is related to $E$ as
\begin{equation}
    A_x(z,t)=-Et + \mbox{const.} + \frac{J}{2N}z^2 + \mathcal{O}(z^4).
    \label{GaugeField}
\end{equation}
Here we have employed the gauge $\partial_x A_t =0$. Thus, the Lagrangian density in the D7-brane action~(\ref{DBIaction}) is explicitly written as
\begin{widetext}
\begin{equation}
    \mathcal{L}_{D7}=-N \cos^3\theta g_{xx} \sqrt{|g_{tt}|g_{xx}g_{zz}-g_{zz}(\dot{A_x})^2+|g_{tt}|(A_{x}')^2},
    \label{D7action}
\end{equation}
where the prime and the dot denote the differentiation with respect to $z$ and $t$, respectively. The induced metric agrees with the metric of AdS-BH~(\ref{AdSBH}) except for $g_{zz}=1/z^2+\theta'(z)^2$. According to the AdS/CFT dictionary, the current density $J$ (in the $x$ direction) is given by $J=\partial \mathcal{L}_{D7} /\partial A_x'$.

Let us perform a Legendre transformation
\begin{eqnarray}
    \tilde{\mathcal{L}}_{D7} &=& \mathcal{L}_{D7}-A_{x}'\frac{\partial \mathcal{L}_{D7}}{\partial A_{x}'} \nonumber \\ 
    &=& -\sqrt{g_{zz}\left( g_{xx}-\frac{E^2}{|g_{tt}|} \right) \left( N^2|g_{tt}|g_{xx}^2\cos^{6}\theta-J^2\right)},
    \label{D7LagrangianLegendre}
\end{eqnarray}
so that $J$ becomes a control parameter. 
\end{widetext}
The Euler-Lagrange equation for $\theta$ is
\begin{equation}
    \frac{\partial}{\partial z}\frac{\partial \tilde{\mathcal{L}}_{D7}}{\partial \theta'}-\frac{\partial \tilde{\mathcal{L}}_{D7}}{\partial \theta}=0.
    \label{EulerLagrange}
\end{equation}
In addition, requiring the on-shell D7-brane action~(\ref{D7LagrangianLegendre}) to be real, we can determine the relationship between $J$ and $E$ as
\begin{equation}
    J=\pi N T(e^2+1)^{1/4} \cos^3 \theta (z_*) E ,
    \label{current}
\end{equation}
where $z_*$ is the point at which $\tilde{\mathcal{L}}_{D7}$ equals zero~\cite{KO_JHEP}. More explicitly, $z_*=(\sqrt{e^2+1}-e)^{1/2} z_H$ and $e=2E/(\pi \sqrt{2 \lambda} T^2)$. 

We need to solve the equation of motion (EOM)~(\ref{EulerLagrange}) numerically in order to obtain $\theta (z_{*})$ explicitly. We employ two boundary conditions: $\theta (z) / z |_{z=0} = m_q$ and $\theta' |_{z=z_*}=[B-\sqrt{B^2+C^2}]/(C z_*)$. Here $B=3z_{H}^8 + 2z_{H}^4 z_{*}^{4}+3z_{*}^{8}$ and $C=3(z_{*}^{8}-z_{H}^{8})\tan\theta (z_{*})$. The second boundary condition is derived from the EOM at $z=z_{*}$~\cite{AFJK_JHEP}. (See also Ref.~\cite{SN_PTP}.) After solving the EOM numerically under these boundary conditions, we pick out the corresponding values of $J$ and $E$ so that $m_q$ agrees with the designed value. Since the numerical analysis becomes unstable at $z=0$, $z=z_H$ and $z=z_*$, we avoid these points by introducing cutoffs. 

\begin{figure}[tb]
\begin{center}
\includegraphics[width=8cm,bb=0 0 400 247,clip]{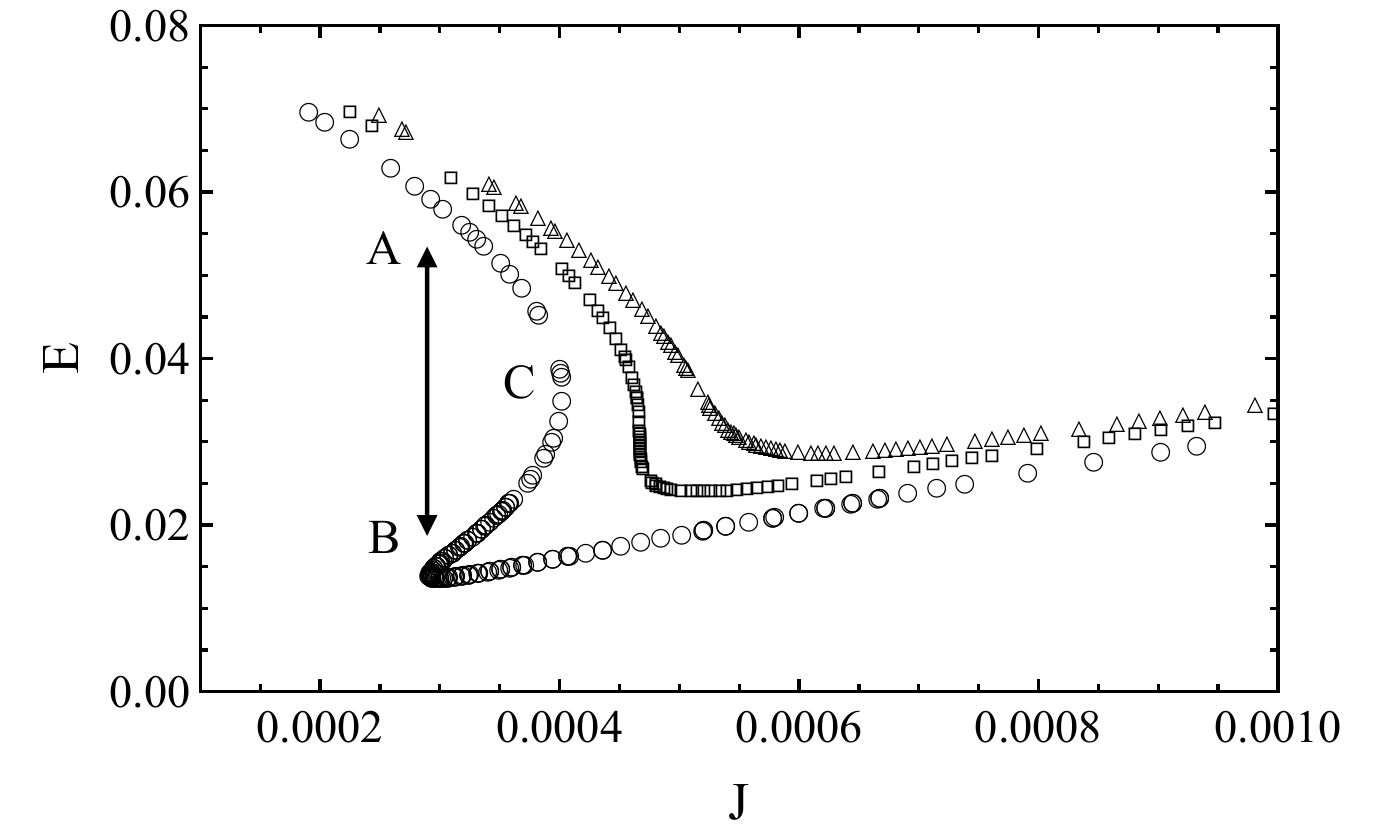}
\caption{The $J$-$E$ characteristics at various temperatures: $T=0.34378>T_c$ (circle), $T=0.34365=T_c$ (box), and $T=0.34356<T_c$ (triangle). }
\label{fig:JE}
\end{center}
\end{figure}

We choose $m_q=1$ and $N=1$ for simplicity. In other words, our $J$ is understood as $J/N$ when we assign general value to $N$. The $J$-$E$ characteristics at various temperatures are shown in Fig.~\ref{fig:JE}. 

For $T<T_c$ the NDC region, where the slope of the $J$-$E$ curve is negative, is smoothly connected to the PDC region, where the slope is positive. On the other hand, for $T>T_c$ there is an intermediate region between the NDC region and the PDC region, where $E$ has three different possible values at a given $J$. Since the value of $E$ has to be selected to one of them, $E$ jumps to another value at some point in this region. It was proposed that the transition point is determined by a thermodynamic potential defined by using the Hamiltonian of the D7-brane~\cite{SN_PRL}. The Hamiltonian density is given by
\begin{eqnarray}
       \tilde{\mathcal{H}}_{D7} &=& \dot{A_x} \frac{\partial \tilde{\mathcal{L}}_{D7}}{\partial \dot{A_x}}-\tilde{\mathcal{L}}_{D7}  \nonumber \\
    &=& g_{xx}\sqrt{|g_{tt}|g_{zz}}\sqrt{\frac{N^2 \cos^6 \theta |g_{tt}|g_{xx}^2 -J^2}{|g_{tt}|g_{xx}-E^2}}.
    \label{HamiltonianDensity}
\end{eqnarray}
Then the thermodynamic potential is defined as
\begin{equation}
    \tilde{F}_{D7} (T, J; m_q)=\lim_{\epsilon \rightarrow 0} \left[ \int_{\epsilon}^{z_H} dz  \tilde{\mathcal{H}}_{D7}-L_{\mbox{\tiny{count}}}(\epsilon) \right],
    \label{Hamiltonian}
\end{equation}
where $L_{\mbox{\tiny{count}}}$ denotes the counterterms that renormalize the divergence at the boundary $z=0$. $L_{\mbox{\tiny{count}}}$ is given by 
\begin{equation}
    L_{\mbox{\tiny{count}}}=L_1+L_2-L_F+L_f,
   \label{Lcount}
\end{equation}
where each term of (\ref{Lcount}) is given in Ref.~\cite{KO_JHEP} as
\begin{eqnarray}
    L_1&=&\frac{1}{4}\sqrt{-\det{\gamma_{ij}}}, \\
    L_2&=&-\frac{1}{2}\sqrt{-\det{\gamma_{ij}}}\theta(\epsilon)^2, \\
    L_f&=&\frac{5}{12}\sqrt{-\det{\gamma_{ij}}}\theta(\epsilon)^4, \\
    L_F&=&\frac{1}{2} E^2\log{\kappa \epsilon}.
\end{eqnarray}
Here $\gamma_{ij}$ is the induced metric on the $z=\epsilon$ slice and $\kappa$ is a factor in order to make the argument of the logarithm dimensionless. The value of $\kappa$ is scheme dependent, and we have chosen this value as one of the possible choices so that $\partial^2 \mathcal{L}_{D7} / \partial E^2 =0$ for vacuum~($T=0,~E=0,~m_q\neq0$). It has been found that the stable state has the lowest $E$ at a given $J$~\cite{SN_PRL}. As a result, the transition point between the NDC phase and the PDC phase is the point indicated by the arrow between A and B in Fig.~\ref{fig:JE}. We call the transition for $T>T_c$ the first-order transition because $E$ changes discontinuously. We call the transition for $T=T_c$ the second-order transition because the differential resistivity $\partial E/\partial J$ diverges there while the $\sigma=J/E$ changes continuously~\cite{SN_PRL}.

\section{Results}
\label{CriticalExponents}
In this section, we consider the critical phenomena of the nonequilibrium phase transition given in the previous section.

\subsection{$\beta$ and $\delta$}
\begin{figure}[tb]
\begin{center}
\includegraphics[width=7cm, bb=0 0 400 247,clip]{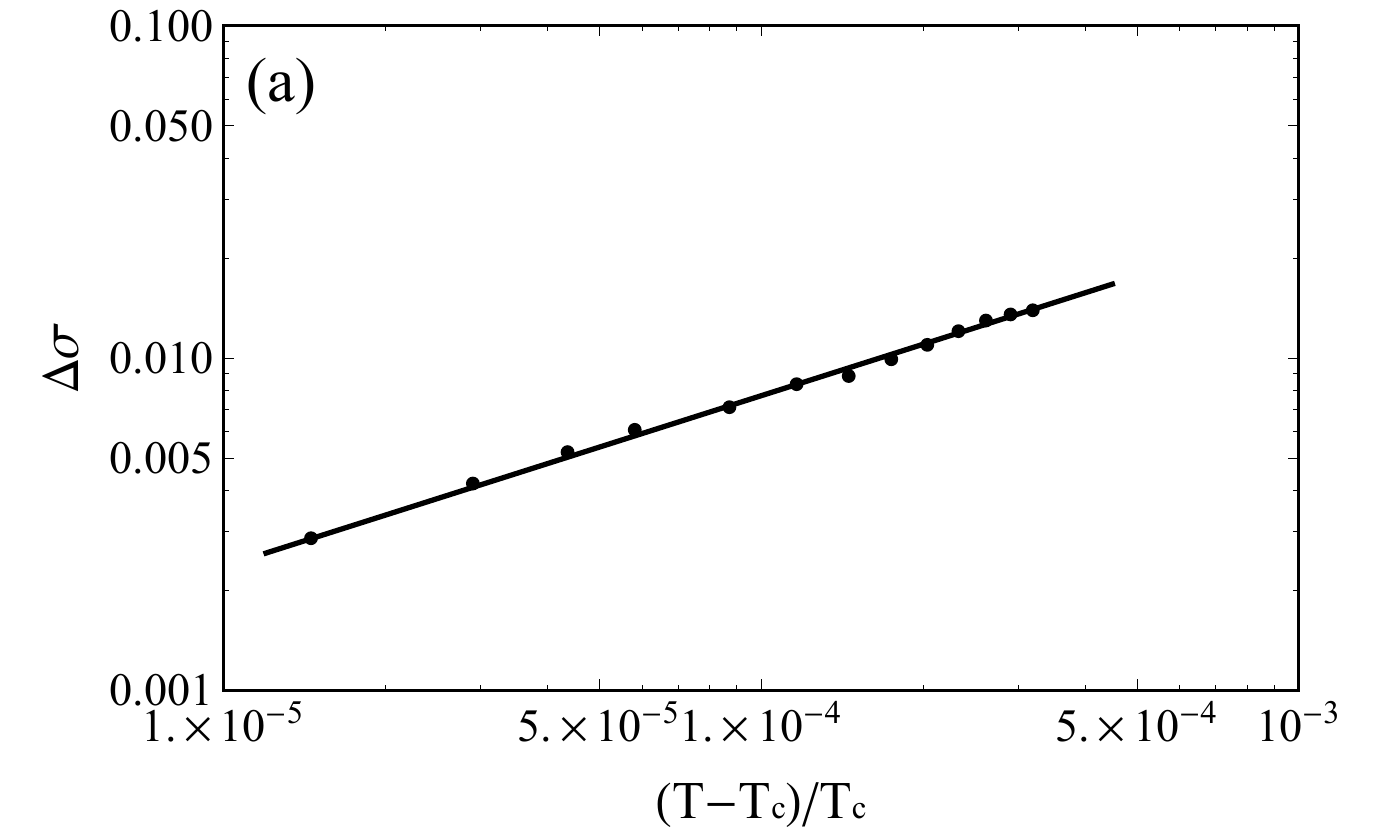}
\includegraphics[width=7cm, bb=0 0 400 247,clip]{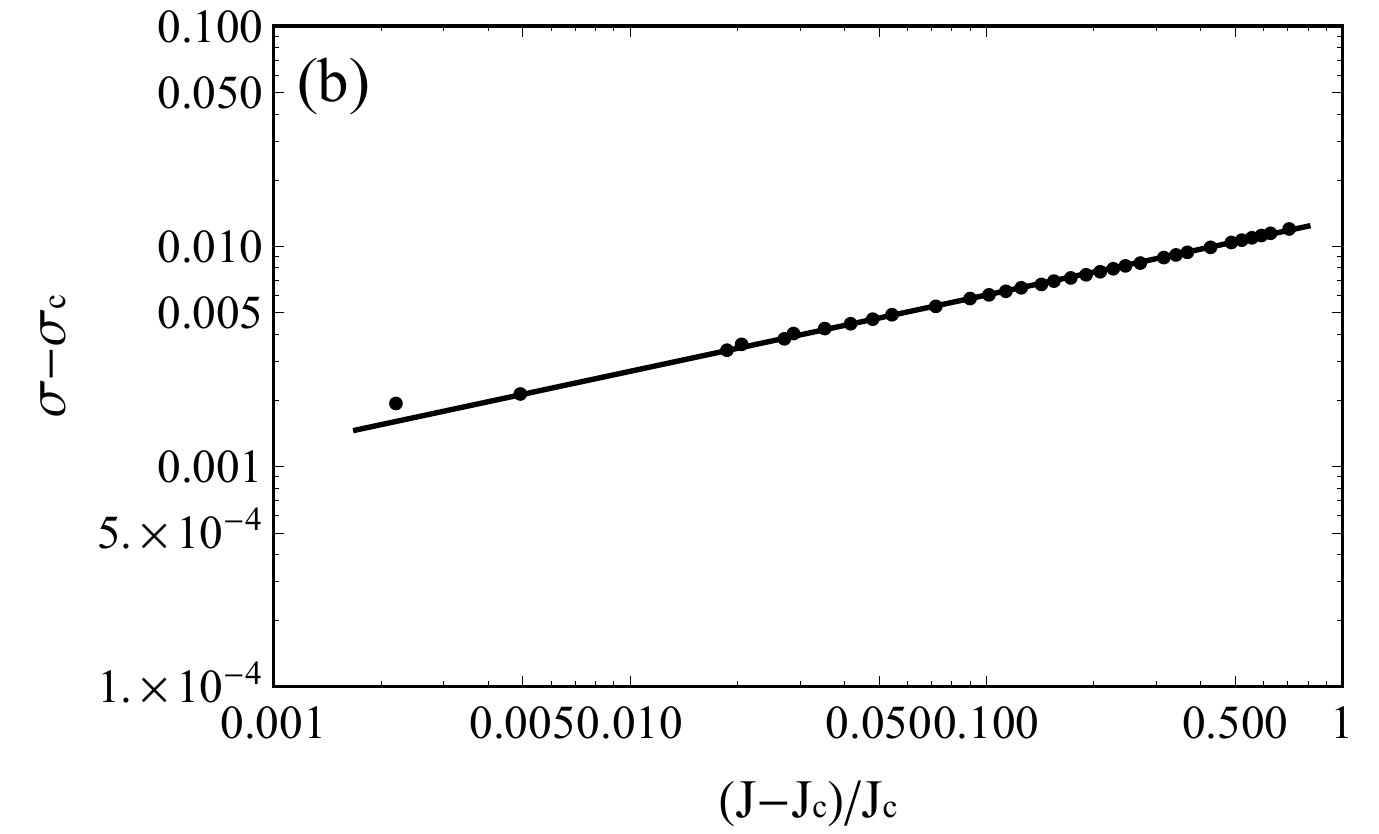}
\caption{(a) Critical behavior of the difference of the conductivity $\Delta \sigma$ near the critical point and (b) that of $\sigma-\sigma_c$.}
\label{fig:betadeltasigma}
\end{center}
\end{figure}

In our nonequilibrium phase transition, the critical exponents $\beta$ and $\delta$ are defined in Ref.~\cite{SN_PRL} as
\begin{equation}
    \Delta \sigma \propto |T-T_{c}|^{\beta}, \;\; |\sigma-\sigma_{c}| \propto |J-J_c|^{1/\delta},
    \label{BetaDelta}
\end{equation}
where $T$ is the heat-bath temperature and $\Delta \sigma$ is the difference of the conductivity between the PDC phase and the NDC phase at a transition point. $\sigma_c$ and $J_c$ are the conductivity and the current density at the critical point, respectively. $\Delta \sigma$ is evaluated along the line of the first-order phase transition. The value of $\delta$ is evaluated along the line of $T=T_c$. These definitions correspond to Eqs.~(\ref{CEbeta}) and~(\ref{CEdelta}) in the Landau theory: the definitions (\ref{BetaDelta}) were proposed by using an assumption that $\sigma-\sigma_c$ and $J-J_c$ play a role of the order parameter and that of the external field, respectively. Our numerical data are shown in Fig.~\ref{fig:betadeltasigma}. We obtain $\beta=0.505 \pm 0.008$ and $\delta=3.008 \pm 0.032$.

\begin{figure}[tb]
\begin{center}
\includegraphics[width=7cm,bb=0 0 397 240,clip]{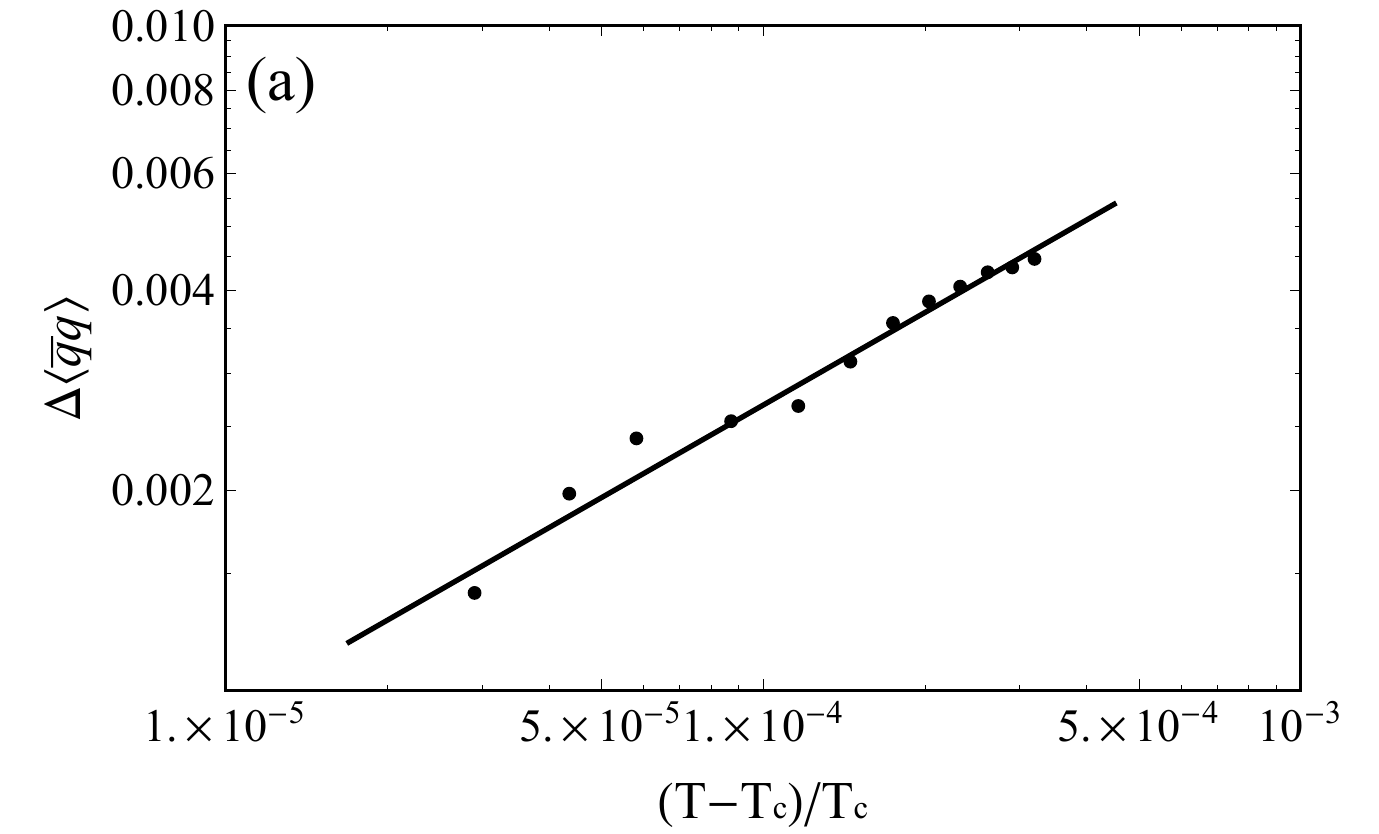}
\includegraphics[width=7cm,bb=0 0 397 240,clip]{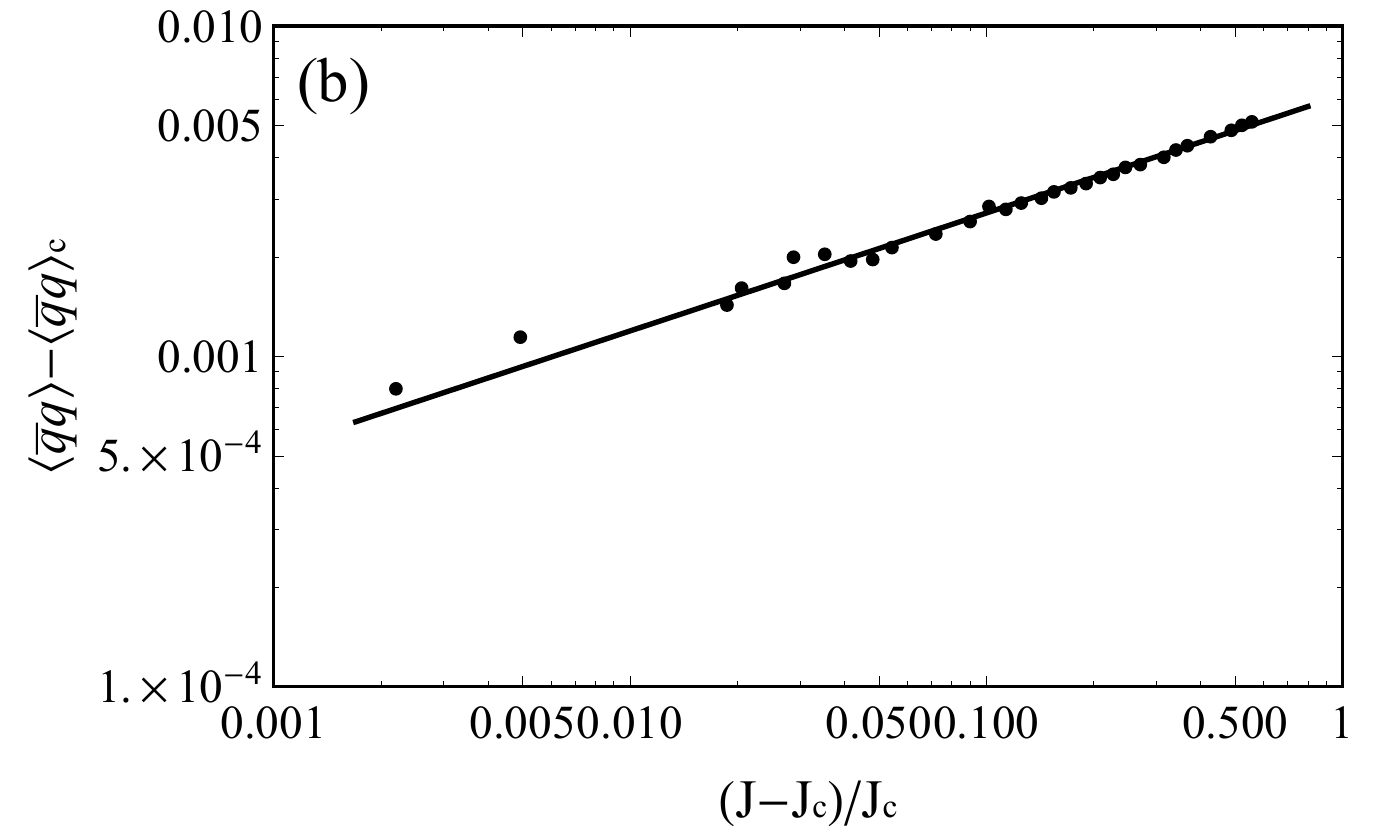}
\caption{(a) Critical behavior of the difference of the chiral condensate $\left< \bar{q} q \right>$ near the critical point and (b) that of $\left< \bar{q}q \right>-\left< \bar{q}q \right>_{c}$.}
\label{fig:betadeltaqq}
\end{center}
\end{figure}

It has been proposed in Ref.~\cite{SN_PRL} that the chiral condensate $\left< \bar{q}q \right>$ is another candidate for the order parameter. Then we have another definition of the critical exponents:
\begin{equation}
    \Delta \left< \bar{q}q \right> \propto |T-T_{c}|^{\beta_{\mbox{\tiny{chiral}}}}, \;\; |\left< \bar{q}q \right>-\left< \bar{q}q \right>_{c}| \propto |J-J_c|^{1/\delta_{\mbox{\tiny{chiral}}}},
    \label{BetaDeltaChiral}
\end{equation}
where $\left< \bar{q}q \right>_{c}$ is $\left< \bar{q}q \right>$ at the critical point. We show the numerical results for chiral condensate in Fig.~\ref{fig:betadeltaqq}. We find that these critical exponents are $\beta_{\mbox{\tiny{chiral}}}=0.515 \pm 0.029$ and $\delta_{\mbox{\tiny{chiral}}}=2.999 \pm 0.061$. We have reconfirmed the results found in Ref.~\cite{SN_PRL}. Note that all of these values agree with those of the Landau theory given in~(\ref{MFCriticalExponents}) within the numerical error.

\subsection{$\gamma$}
This section is the main part of the present paper which is about the definition and calculation of the critical exponent $\gamma$. First we define the critical exponent $\gamma$ for our nonequilibrium phase transition. In Sec.~\ref{LandauTheory}, we have reviewed that the critical exponent $\gamma$ in the Landau theory is defined by using the magnetic susceptibility $\chi=\partial M / \partial H$, where $M$ is the magnetization and $H$ is the external magnetic field. Near the critical point, the magnetic susceptibility behaves as $\chi \propto |T-T_c|^{-\gamma}$. In our nonequilibrium phase transition, since we use either the conductivity or the chiral condensate as the order parameter, it is natural to generalize the definition of $\chi$ as
\begin{equation}
    \tilde{\chi}=\frac{\partial (\sigma - \sigma_{c})}{\partial J}, \;\; \tilde{\chi}_{\mbox{\tiny{chiral}}}=\frac{\partial ( \left< \bar{q}q \right>-\left< \bar{q}q \right>_{c})}{\partial J},
    \label{ChiDef}
\end{equation}
where $J$ is again assumed to act as the external field.
We can rewrite $\tilde{\chi}$ by using the definition of conductivity $\sigma=J/E$, 
\begin{equation}
    \tilde{\chi}=\frac{1}{E}-\frac{J}{E^2}\frac{\partial E}{\partial J},
    \label{Chi}
\end{equation}
so that it can be calculated from the $J$-$E$ characteristics. 

\begin{figure}[tb]
\begin{center}
\includegraphics[width=7cm,bb=0 0 397 240,clip]{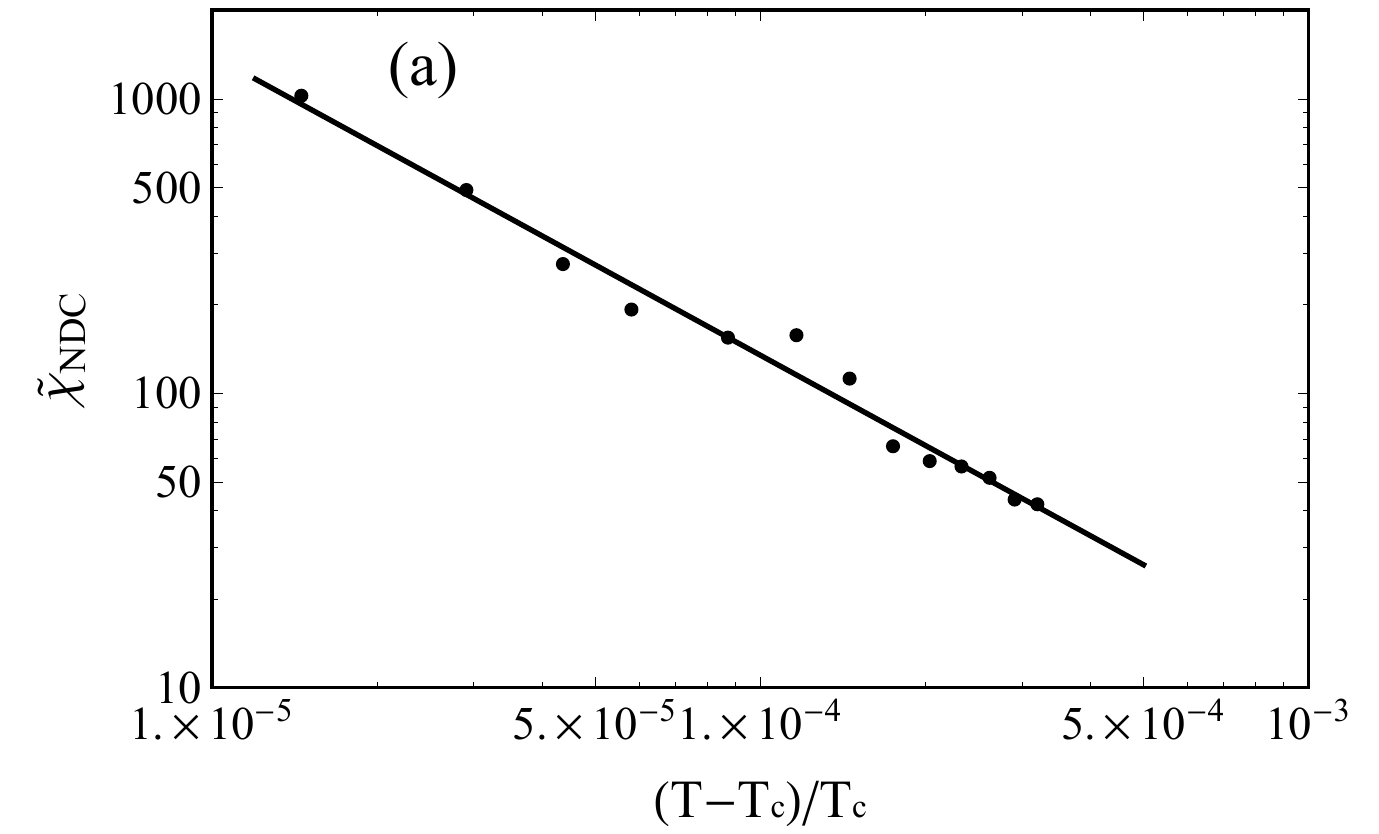}
\includegraphics[width=7cm,bb=0 0 397 240,clip]{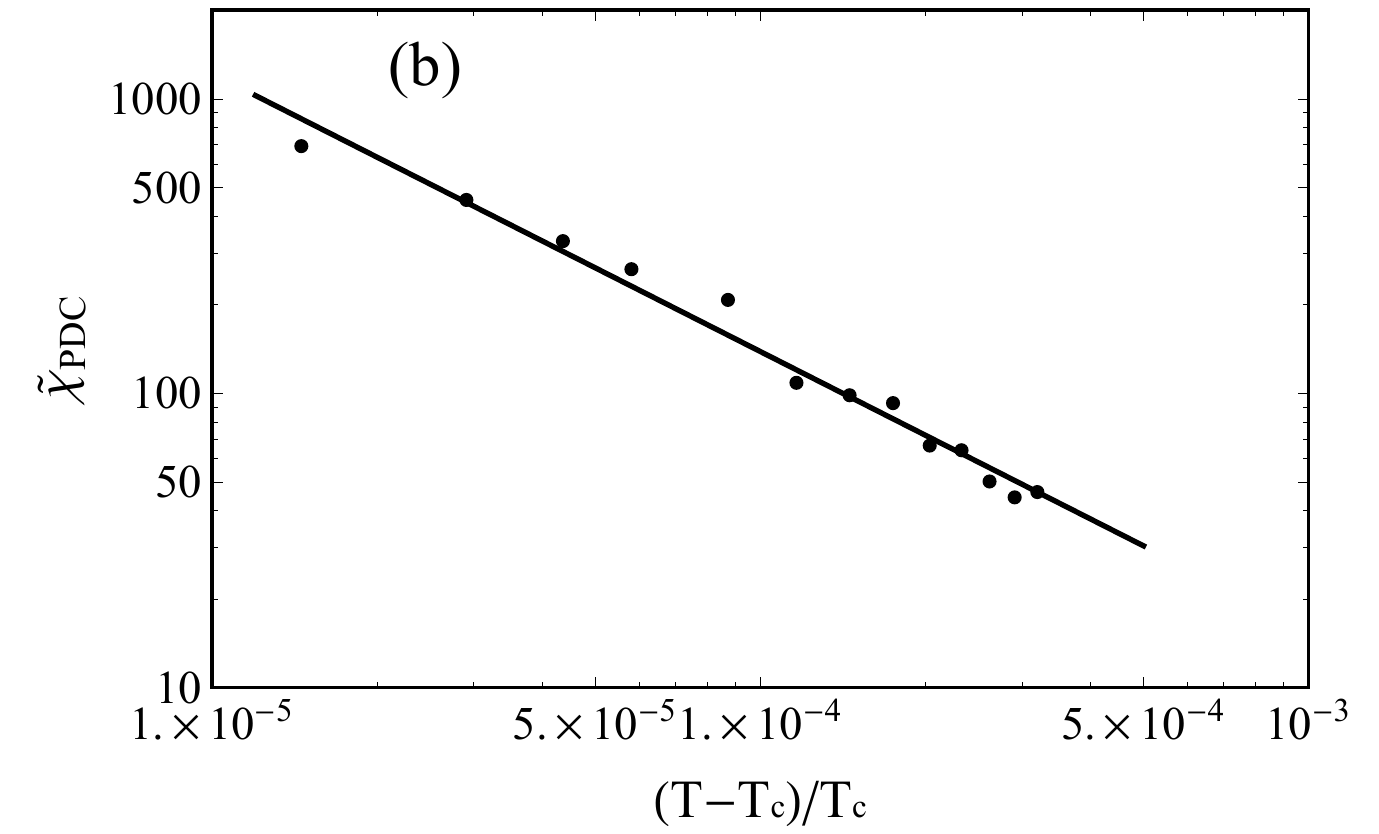}
\caption{Critical behaviors of $\tilde{\chi}$ for $T>T_c$ (a) in the NDC phase and (b) in the PDC phase.}
\label{fig:gammasigma}
\end{center}
\end{figure}

We propose to define the critical exponent $\gamma$ as
\begin{equation}
    \tilde{\chi} \propto |T-T_{c}|^{-\gamma}
    \label{Gamma}
\end{equation}
in our nonequilibrium phase transition.\footnote{
Note that if the state with larger $E$ were more stable, the transition point would be at C in Fig.~\ref{fig:JE}. However, we cannot calculate $\tilde{\chi}$ at this point because $\partial E/\partial J$ is always divergent there.}
There are two possible definitions of $\tilde{\chi}$ for $T > T_c$: that in the NDC phase and that in the PDC phase
\begin{equation}
    \tilde{\chi}_{\mbox{\tiny{NDC}}}=\frac{\partial (\sigma_{\mbox{\tiny{NDC}}}-\sigma_{c})}{\partial J}, \;\; \tilde{\chi}_{\mbox{\tiny{PDC}}}=\frac{\partial (\sigma_{\mbox{\tiny{PDC}}}-\sigma_{c})}{\partial J}.
    \label{ChiDef2}
\end{equation}
As shown in Fig.~\ref{fig:gammasigma}, the behaviors of the susceptibilities in these phases are similar to each other and it is found that each value of $\gamma$ is $\gamma_{\mbox{\tiny{NDC}}}=1.018 \pm 0.043$ and $\gamma_{\mbox{\tiny{PDC}}}=1.014 \pm 0.042$. We find that they agree with that from the Landau theory~(\ref{MFCriticalExponents}), $\gamma=1$, within the numerical error.

\begin{figure}[tb]
\begin{center}
\includegraphics[width=8cm,bb=0 0 400 270,clip]{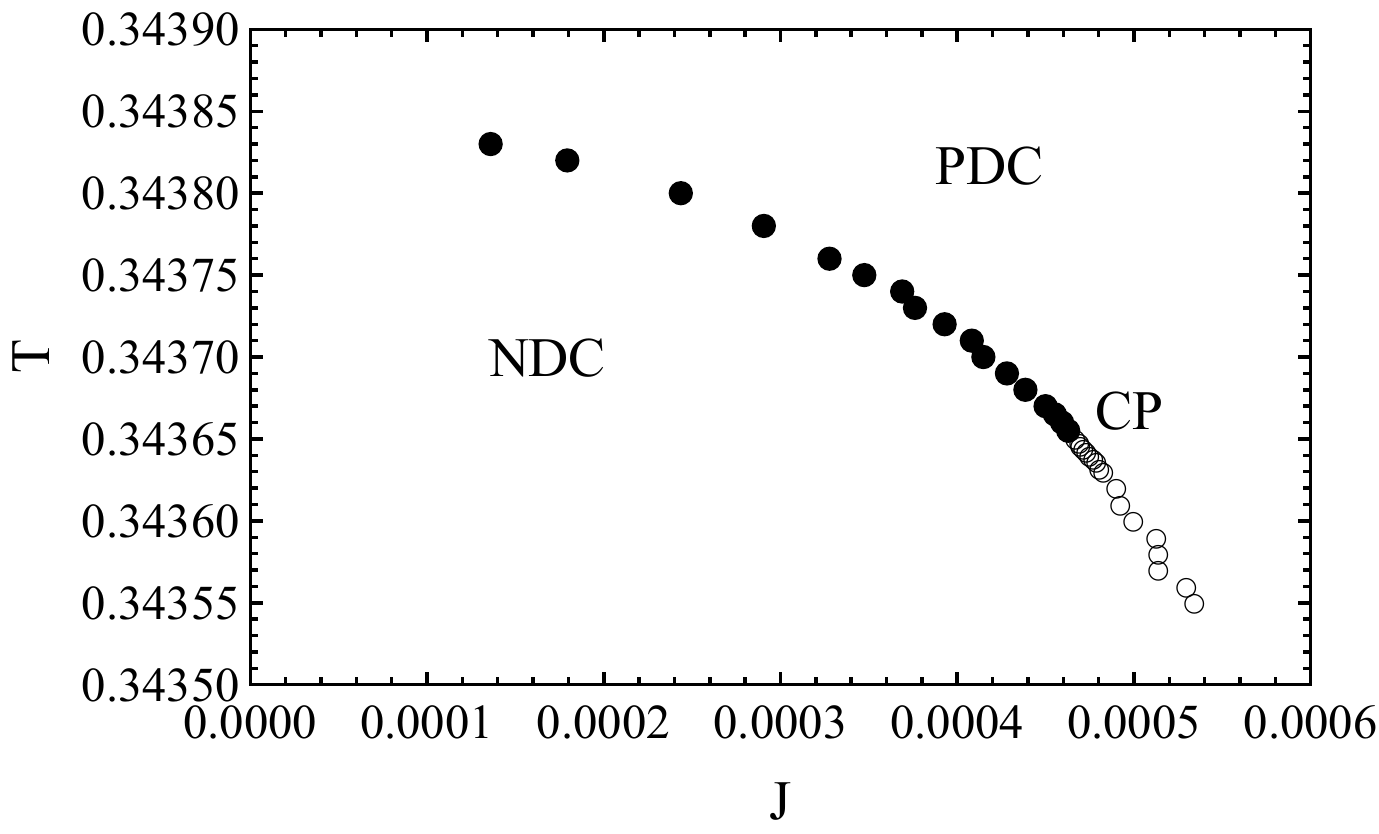}
\caption{The phase diagram for our nonequilibrium phase transition. The filled circles are on the line of the first-order phase transition. The critical point (CP) is at $T_c=0.34365$. The open circles are the inflection points where $\sigma=\sigma_c=0.0156$.}
\label{fig:phasediagram}
\end{center} 
\end{figure}

In addition, we evaluate the $\gamma$ for $T<T_c$. In the liquid-vapor phase transition, the susceptibility should be calculated along the critical isochore in the crossover region. Therefore, it is necessary to determine the line that corresponds to the critical isochore for our nonequilibrium phase transition. In analogy with the ferromagnet phase transition or the liquid-vapor transition, we choose this point as the inflection point in the $J$-$\sigma$ curve. The phase diagram is shown in Fig.~\ref{fig:phasediagram} and it is found that the inflection point for $T<T_c$ is nearly constant with $\sigma=\sigma_c=0.0156$. 

\begin{figure}[tb]
\begin{center}
\includegraphics[width=7cm,bb=0 0 400 270,clip]{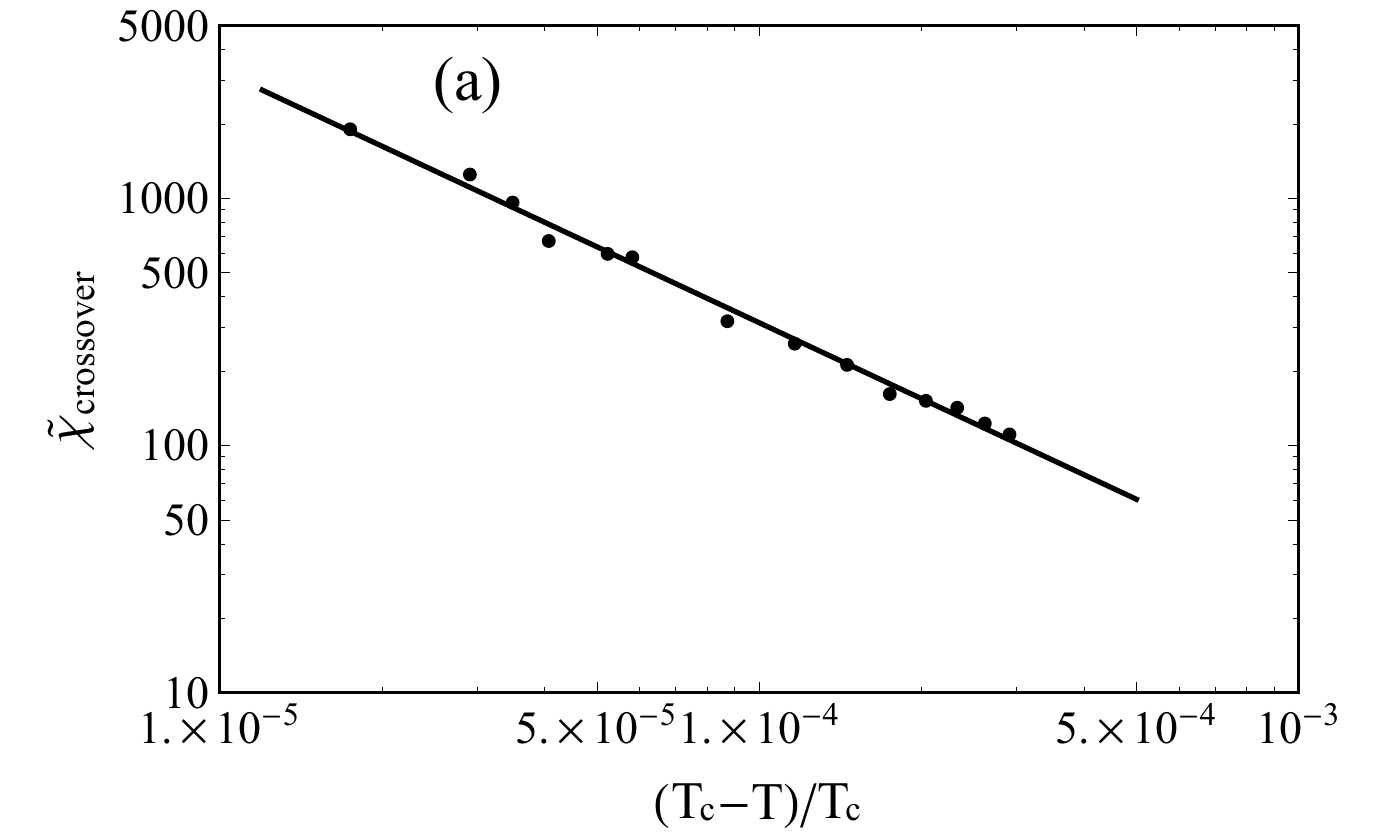}
\includegraphics[width=7cm,bb=0 0 400 270,clip]{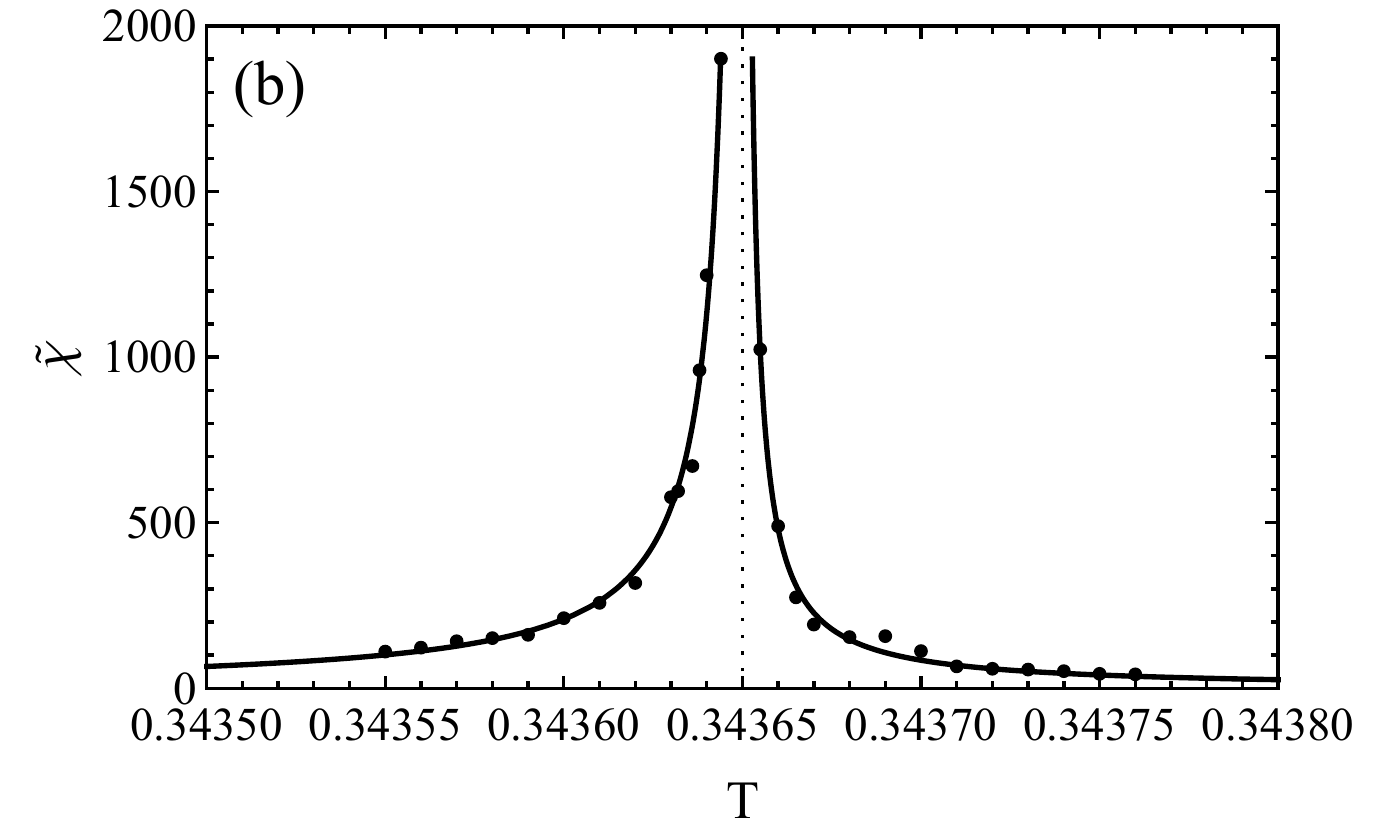}
\caption{(a) Critical behavior of $\tilde{\chi}$ for $T<T_c$ and (b) the divergence of $\tilde{\chi}$ near the critical point.}
\label{fig:gammaacrosssigma}
\end{center}
\end{figure}

We show the relationship between the values of $\tilde{\chi}$ and the temperature along the $\sigma=\sigma_c$ line in Fig.~\ref{fig:gammaacrosssigma}. The numerical data gives $\gamma_{\mbox{\tiny{crossover}}}=1.022 \pm 0.025$. Furthermore, if we assume that $\gamma_{\mbox{\tiny{NDC}}}=\gamma_{\mbox{\tiny{PDC}}}=\gamma_{\mbox{\tiny{crossover}}}$, we find $\tilde{\chi}_{\mbox{\tiny{crossover}}} / \tilde{\chi}_{\mbox{\tiny{NDC}}}=2.2 \pm 0.4$ and $\tilde{\chi}_{\mbox{\tiny{crossover}}} / \tilde{\chi}_{\mbox{\tiny{PDC}}}=2.0 \pm 0.4$. These results agree with the fact that the critical amplitude ratio in the Landau theory is 2. The critical phenomena are exhibited in Fig.~\ref{fig:gammaacrosssigma}.

\begin{figure}[tb]
\begin{center}
\includegraphics[width=7cm,bb=0 0 397 270,clip]{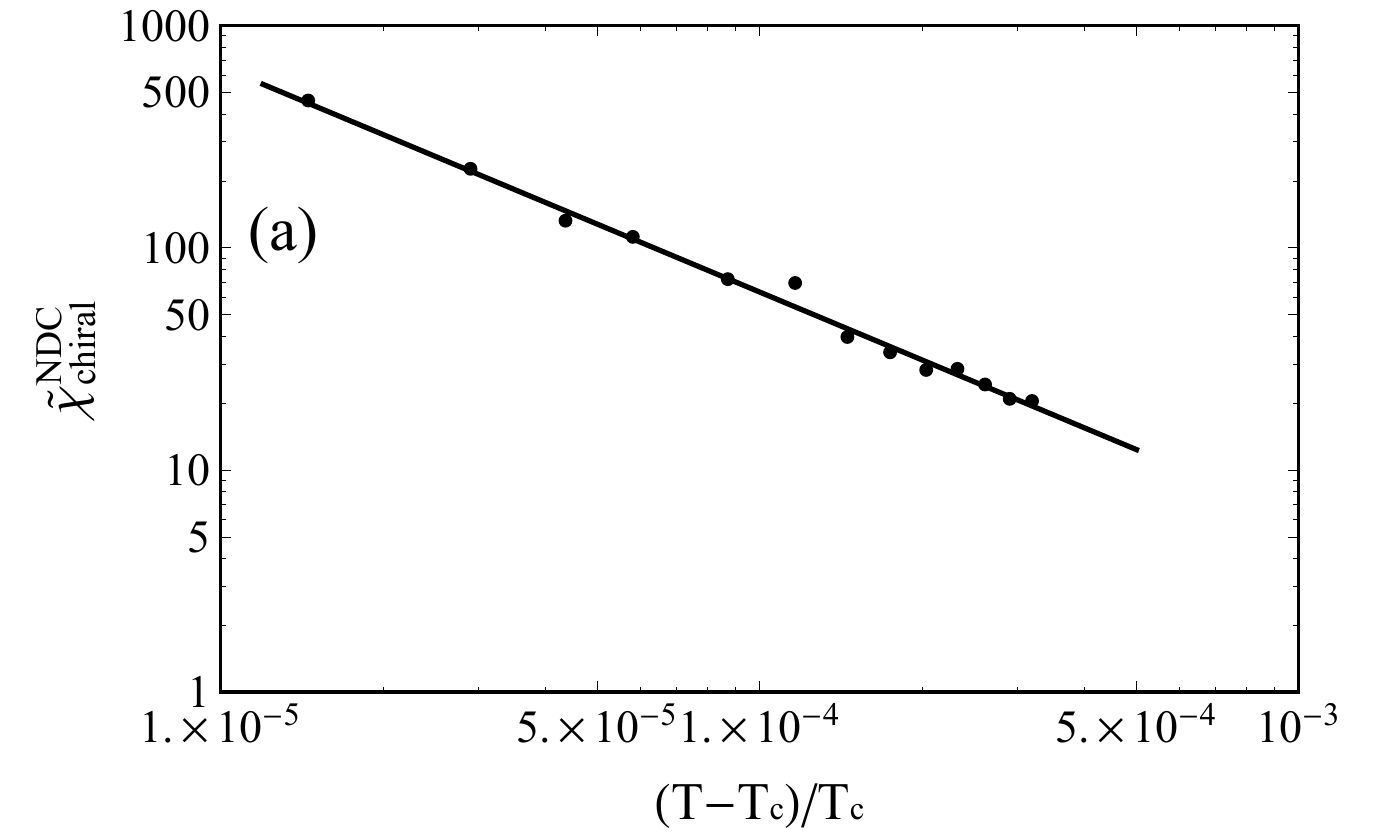}
\includegraphics[width=7cm,bb=0 0 397 270,clip]{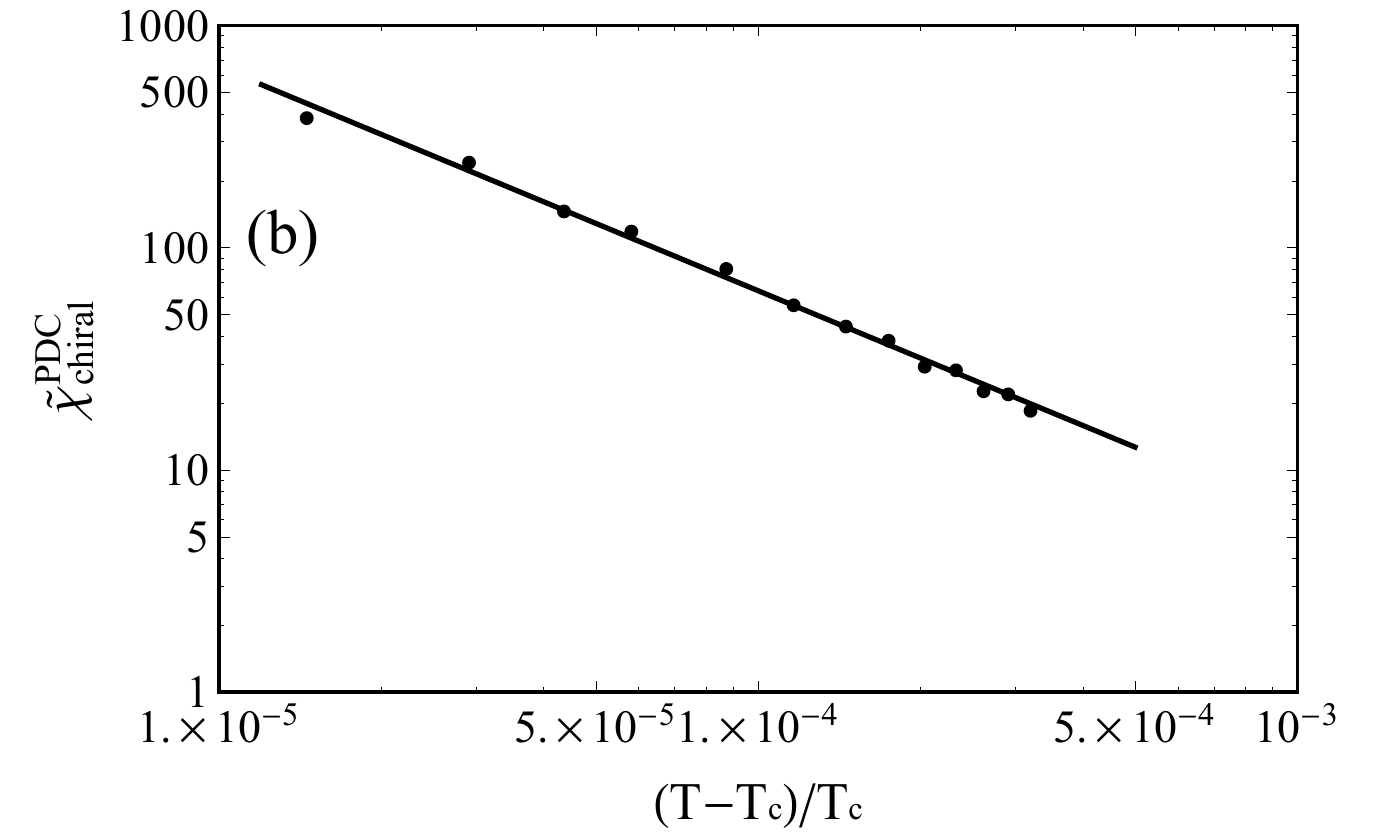}
\includegraphics[width=7cm,bb=0 0 397 270,clip]{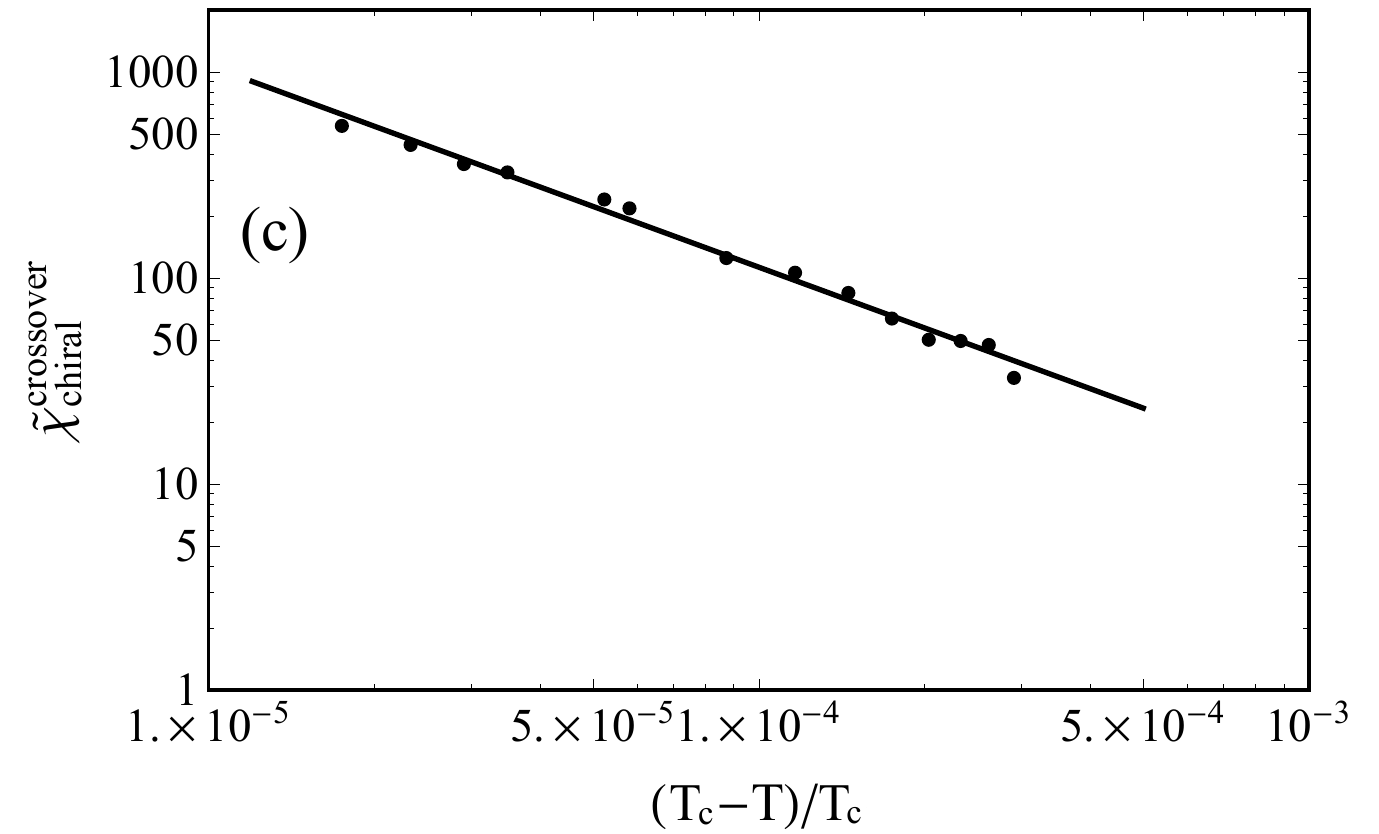}
\caption{Critical behaviors of $\tilde{\chi}_{\mbox{\tiny{chiral}}}$ (a) in the NDC phase, (b) in the PDC phase, and (c) in the crossover region.}
\label{fig:gammaqq}
\end{center}
\end{figure}

All of the above arguments go along with the chiral condensate instead of the conductivity. The obtained values of the corresponding critical exponents are $\gamma_{\mbox{\tiny{chiral}}}^{\mbox{\tiny{NDC}}} =1.015 \pm 0.028$, $\gamma_{\mbox{\tiny{chiral}}}^{\mbox{\tiny{PDC}}} =1.007 \pm 0.022$, and $\gamma_{\mbox{\tiny{chiral}}}^{\mbox{\tiny{crossover}}} = 0.979 \pm 0.029$. They agree with~(\ref{MFCriticalExponents}), again. The corresponding critical amplitude ratios are $\tilde{\chi}_{\mbox{\tiny{chiral}}}^{\mbox{\tiny{crossover}}} / \tilde{\chi}_{\mbox{\tiny{chiral}}}^{\mbox{\tiny{NDC}}}=2.0 \pm 0.3$ and $\tilde{\chi}_{\mbox{\tiny{chiral}}}^{\mbox{\tiny{crossover}}} / \tilde{\chi}_{\mbox{\tiny{chiral}}}^{\mbox{\tiny{PDC}}}=1.9 \pm 0.3$ which agree with 2 within the numerical error.

\section{Conclusion and Discussion}
\label{Conclusion}
We found that the critical exponents of our nonequilibrium phase transition agree with those in the Landau theory: $\beta=1/2$, $\delta=3$, and $\gamma=1$. The critical amplitude ratio of $\tilde{\chi}$ also agreed with that of the Landau theory. Our results satisfy the scaling laws such as the Widom scaling, $\gamma=\beta(\delta-1)$, within the numerical error.

\begin{figure}[tb]
\begin{center}
\includegraphics[width=7cm,bb=0 0 400 247,clip]{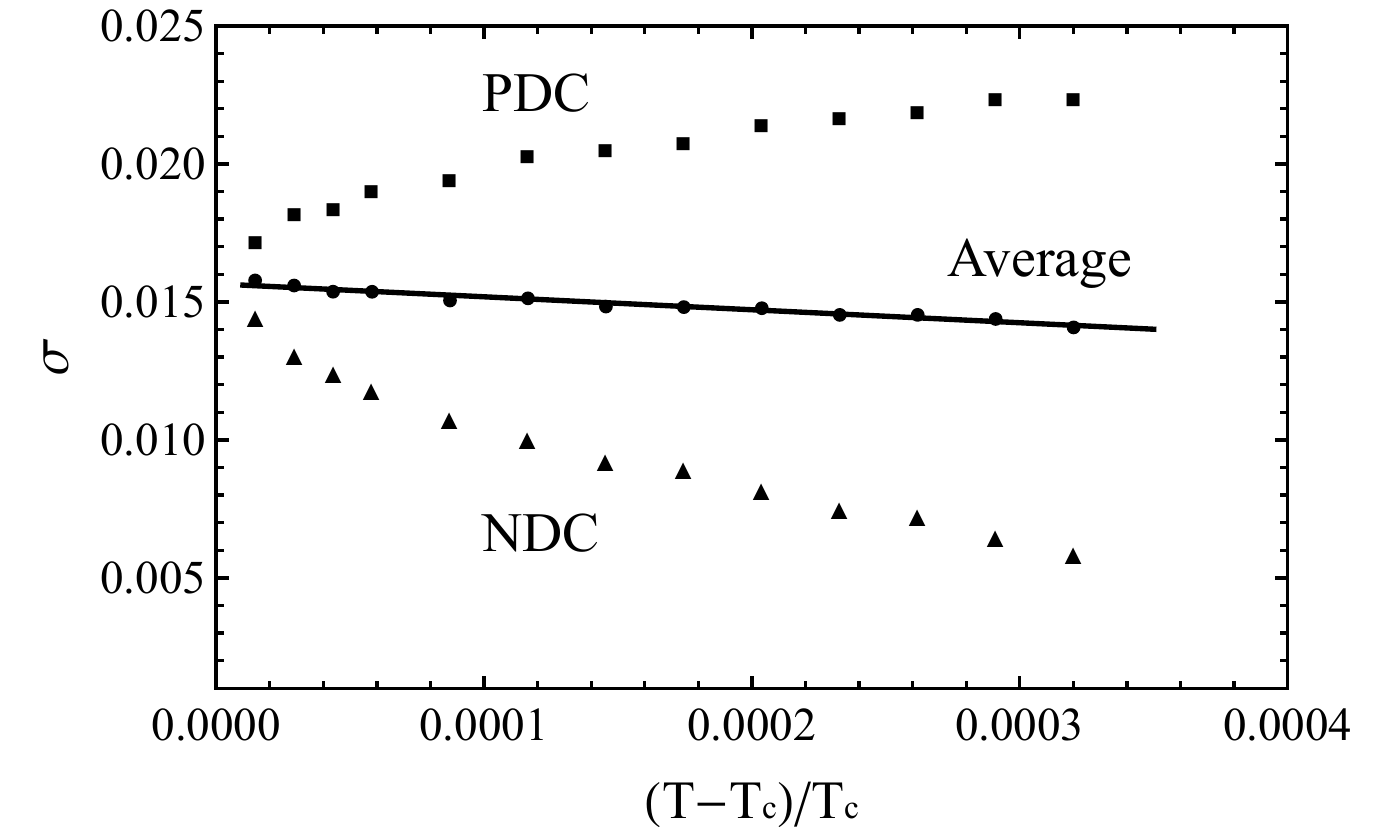}
\caption{Critical behaviors of the conductivities in the NDC phase and the PDC phase and the average of them.}
\label{fig:alphasigma}
\end{center}
\end{figure}

We have two remarks.
There are models of equilibrium phase transitions in which a deviation of the law of rectilinear diameter gives the critical exponent $\alpha$~\cite{WidomRowlinson, Mermin, HemmerStell}.
Let us see how it goes for our case. If we assume that the foregoing method is valid in our system, we may define $\alpha$ as
\begin{equation}
    \sigma_{\mbox{\tiny{ave}}} =\sigma_{c} + A |T-T_{c}|^{1-\alpha},
    \label{Alpha}
\end{equation}
where $A$ is a constant, $\sigma_{\mbox{\tiny{ave}}}=(\sigma_{\mbox{\tiny{NDC}}}+\sigma_{\mbox{\tiny{PDC}}})/2$ and $\sigma_{c}$ is the critical conductivity at $T=T_{c}$. In Fig.~\ref{fig:alphasigma}, we show the conductivities in the PDC phase, those in the NDC phase, and their averages. We obtain the value of the exponent $\alpha=0.048 \pm 0.111$, which agrees with (\ref{MFCriticalExponents}) of the Landau theory. We may define $\alpha_{\mbox{\tiny{chiral}}}$ as
\begin{equation}
    \left< \bar{q}q \right>_{\mbox{\tiny{ave}}}=\left< \bar{q}q \right>_{c}+B|T-T_{c}|^{1-\alpha _{\mbox{\tiny{chiral}}}},
    \label{AlphaChiral}
\end{equation}
where $B$ is a constant,  $\left< \bar{q}q \right>_{\mbox{\tiny{ave}}}= \left( \left< \bar{q}q \right>_{\mbox{\tiny{NDC}}}+\left< \bar{q}q \right>_{\mbox{\tiny{PDC}}} \right) /2$ and $\left< \bar{q}q \right>_{c}$ is the critical value of the chiral condensate.
However, our numerical data shows that $B \simeq 0$: the values of the chiral condensates in each phase are arranged symmetrically with respect to the critical value, as is the case with the ferromagnet phase transition. For this reason, we cannot determine the value of $\alpha_{\mbox{\tiny{chiral}}}$ accurately in this manner. 
We leave more concrete definition of the critical exponent $\alpha$ for our phase transition to future work.\footnote{Note that it is not straightforward to define $\alpha$ by using the heat capacity, since the notion of the heat capacity in NESS is not clear.} 

The second remark is on the relationship with the Landau theory.
In our definitions of the critical exponents, we assumed that $J-J_{c}$ plays a role of the magnetic field $H$ in (\ref{LandauFreeEnergyH}). 
We find that the critical exponent and the critical amplitude ratio of the susceptibility we defined agree with those of the Landau theory of equilibrium phase transitions. Together with the results for $\beta$ and $\delta$, our results state that the critical phenomena in the nonequilibrium phase transition in question have remarkable similarity with those in the Landau theory of equilibrium phase transition.

Coming back to the questions raised in Sec.~\ref{Introduction}, we obtained the answers to the questions 1) and 2) as far as for the nonequilibrium phase transitions considered in this paper: we can define the susceptibility associated with $J$ in a completely parallel manner to that in the Landau theory, and the susceptibility shows critical phenomena with $\gamma=1$. For the question 3), our results suggest that the critical exponents $\gamma$ and $\delta$ associated with $J$ may be formulated by using a theory similar to the Landau theory. However, further investigation is necessary to get the complete answer. This is an issue for future research to explore.\\

\begin{acknowledgments}
The authors are grateful to Y. Fukazawa, T. Hayata, H. Hoshino, S. Kinoshita, Shang-Yu Wu and R. Yoshii for helpful discussions and comments. The work of S. N. was supported in part by JSPS KAKENHI Grant No. JP16H00810, and the Chuo University Personal Research Grant. The work of M. M. was supported by the Research Assistant Fellowship of Chuo University.
\end{acknowledgments}

\end{document}